\documentclass{PoS}
\usepackage{epsf}
\usepackage{cite}
\usepackage{amsmath}
\usepackage{nicefrac}
\usepackage{esint}

\DeclareMathAlphabet{\mathantt}{OML}{antt}{l}{it}
\DeclareMathAlphabet{\mathpzc}{OT1}{pzc}{m}{n}

\def\beq{\begin{equation}}
\def\eeq{\end{equation}}
\def\bea{\begin{eqnarray}}
\def\eea{\end{eqnarray}}
\def\beqa{\begin{equation}\begin{array}{l}}
\def\eeqa{\end{array}\end{equation}}
\def\eqlab#1{\label{eq:#1}}
\def\figlab#1{\label{fig:#1}}

\def\eref#1{(\ref{eq:#1})}
\def\Eqref#1{Eq.~(\ref{eq:#1})}

\def\Figref#1{Fig.~\ref{fig:#1}}

\def\half{\mbox{\small{$\frac{1}{2}$}}}

\def\quarter{\mbox{\small{$\frac{1}{4}$}}}

\def\ga{\gamma} 
\def\de{\delta}

\def\la{\lambda}

\def\si{\sigma}

\def\dd{{\rm d}}

\def\ra{\rightarrow}
\def\nn{\nonumber}

\def\re{\mbox{Re}}
\def\im{\mbox{Im}}

\def\rk{\mathantt{k}}
\def\rr{\mathantt{r}}
\def\ra{\mathantt{a}}

\def\rB{\mathantt{B}}

\DeclareMathOperator\arctanh{arctanh}
\DeclareMathOperator\arccoth{arccoth}
\DeclareMathOperator\arccosh{arccosh}
\DeclareMathOperator\arcsinh{arcsinh}

\title{Surpassing Wigner's causality bound in 
relativistic scattering with zero-range interaction }

\ShortTitle{Surpassing Wigner's causality bound}

\author{Vladimir Pascalutsa\thanks{Supported by the Deutsche Forschungsgemeinschaft (DFG) through Collaborative 
Research Center SFB 1044.}
       \\
       Institut f\"ur Kernphysik, Johannes Gutenberg Universit\"at Mainz,  D-55099 Mainz, Germany
       }

\abstract{It is shown that the relativistic zero-range potential scattering surpasses
Wigner's causality bound while being consistent with causality.
The relativistic theory shows in addition a richer analytic structure
such as a $K$-matrix pole necessarily accompanying the bound-state solution.
Possible implications of these results
for the effective-field theory of nuclear forces are briefly considered.}

\FullConference{The 8th International Workshop on Chiral Dynamics, CD2015\\
		29 June 2015 -- 03 July 2015\\
		Pisa, Italy }

\begin{document}

\section{Introduction}
\label{sec1}

Wigner's seminal work on 
causality bounds for the effective range of low-energy scattering~\cite{Wigner:1955zz}
has been revisited quite recently in connection to the effective-field-theoretic (EFT)
description of few-nucleon systems and cold atoms, 
see e.g.~\cite{Phillips:1996ae,PavonValderrama:2005wv,Hammer:2009zh,Elhatisari:2012ym}. 
Zero-range forces play an important role in these considerations as
they are expected to provide a leading-order description of any finite-range force,
be it nuclear or Van der Waals. Indeed, the very low-energy (long-distance) probes of systems bound
by finite-range forces cannot resolve the extent at which the forces act, and hence 
the zero-range approximation should naively be fine. For the nuclear force, however, it does not
appear to be too fine. As first noted by Phillips and Cohen~\cite{Phillips:1996ae}, in
case of zero-range forces the Wigner's causality bound infers negative values
for the $s$-wave effective-range parameters, in appreciable 
disagreement with what is observed
in nucleon-nucleon ($NN$) scattering. This problem can be overcome by treating range corrections in perturbation theory, along with other interactions needed for renormalization-group invariance~\cite{vanKolck:1998bw,Birse:1998dk,Gegelia:1998gn}.
 Further
difficulties arise, however, when pions are included (perturbatively) in this framework, see e.g.~\cite{Cohen:1999iaa} and references therein.
A commonly accepted solution nowadays is to
``promote" a finite-range (one-pion-exchange) force  into the leading order,
see Refs.~\cite{vanKolck:1999mw,Bogner:2003wn,Epelbaum:2005pn} for reviews. 
Here, however, we would like to pursue a different route and demonstrate
that a relativistic theory of zero-range forces can both be
consistent with causality and yield positive effective-range parameters. 

More specifically, introducing the $s$-wave scattering phase-shift $\de(\rk)$,
which is a function of the relative momentum $\rk$, 
the effective-range expansion is written as:
\beq
\eqlab{ERE}
 \rk \cot \de(\rk)  = -\frac{1}{\ra} 
+ \frac{1}{2}  \sum_{n=1}^\infty (-1)^{n+1} \, \rr_n \, \rk^{2n},
\eeq
where $\ra$ is the scattering length, $\rr_1$ is the effective range,
and $\rr_{n\geq 2}$ are (up to  an overall factor of 2) the effective-shape parameters. 
While Wigner's causality bound for scattering through a $\de$-function
potential (zero-range force)
yields \cite{Phillips:1996ae}:
\beq
\rr_1 \leq 0 \quad\mbox{(Wigner's bound)},
\eqlab{Wignerbound}
\eeq  
we establish here that the effective range is non-negative for causal scattering, together in fact
with all the effective-shape parameters, i.e.:
\beq
\rr_n \geq 0\quad \mbox{(present work)}.
\eqlab{newbound}
\eeq
This result is in near perfect disagreement with Wigner's bound,
however, will be shown to reconcile with it in the non-relativistic limit
where $\rr_1=0$. Away from non-relativistic limit this result
may open up a venue for an EFT description of nuclear forces where
the pion exchange is suppressed with respect to the zero-range
interaction. 

\section{Light-by-light sum rule as causality criterion}
Given the nearly perfect disparity of the two causality bounds quoted above, 
we start by noting that they are based
on different interpretations of causality. Wigner's bound is based on positivity
of time delay between the incoming and scattered wave, which translates
into the following condition for the phase shift \cite{Nussenzveig:1972}:
\beq
\dd \de/\dd \rk \geq (2\rk)^{-1} \sin2\de.
\eeq
Taking here $\rk\to 0$ one arrives to \Eqref{Wignerbound}.
We, on the other hand, adopt a causality criterion based
on dispersion theory. Nameley, we follow up on the proposal~\cite{Pascalutsa:2011sr} to exploit the analog of 
the Gerasimov-Drell-Hearn (GDH) sum rule for the light-light ($\ga\ga$) system~\cite{Gerasimov:1973ja,Brodsky:1995fj,Pascalutsa:2010sj}:
\beq
\int_{0}^\infty\!\! \dd s\,  \frac{\si_2(s)-\si_0(s)}{s}=0,
\eqlab{HDsr}
\eeq
where $\si_2(s)$ and $\si_0(s)$ are the cross sections of two-photon fusion process ($\gamma \gamma \to X$) with photons circularly polarised in the same or opposite directions, respectively. The total invariant energy squared is
$s=(q_1+q_2)^2$, for $q_1$ and $q_2$ the colliding photon four-momenta.

The validity of this sum rule relies on only general principles such
as Lorentz and gauge symmetries, unitarity and analyticity. The latter
requirement is associated with causality and is the less trivial to satisfy
in a given modeling of these cross sections. This is why the sum rule verification is an indicator of causality above all the other
aforementioned principles.  
\begin{figure}
\centering
  \includegraphics[width=16cm]{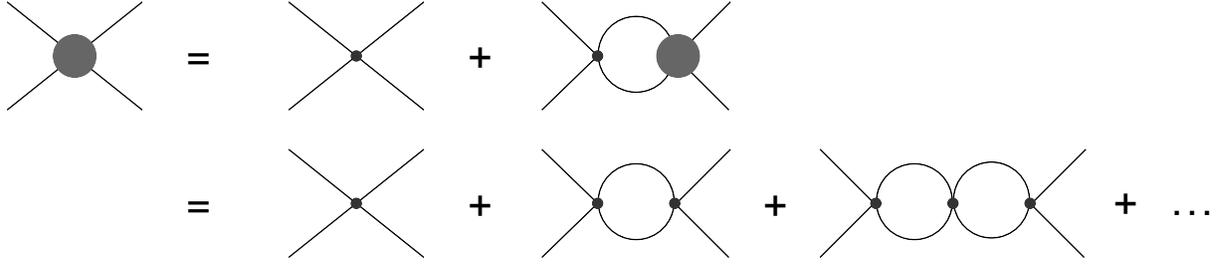}\\
  \caption{The Bethe-Salpeter equation and its iterative solution.}
    \figlab{BSE}
\end{figure}

The sum-rule criterion is applicable to a relativistic scattering theory
by constructing a particle-antiparticle scattering amplitude and 
then considering the $\ga\ga$ fusion into the pair. Assuming, for instance,
that the scattering amplitude is found as the solution of the Bethe-Salpeter equation graphically represented
in \Figref{BSE}, the corresponding $\ga\ga$-fusion process is 
given by diagrams in \Figref{LbL}.

\begin{figure}[hb]
\centering
  \includegraphics[width=9cm]{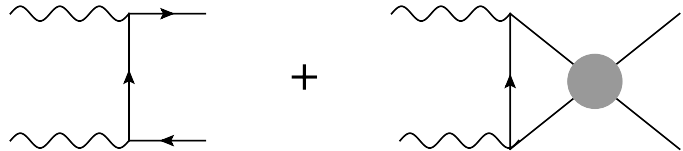}\\
  \caption{Photon-photon fusion with rescattering.}
    \figlab{LbL}
\end{figure}

For the relevant
case of the $\de$-function potential, given in momentum space
by a constant $V=\la$,
this criterion has first been employed
by Pauk~{\it et al.}~\cite{Pauk:2013hxa}, who showed that the
sum rule is satisfied, unless $\la\in (-8\pi^2, 0)$.

We recall that
the solution of the Bethe-Salpeter equation ($T=V+VGT$) 
is algebraic in this case and for the equal-mass situation reads:\footnote{Although this solution may seem arbitrary from field-theoretic point of view, it emerges in the $O(N)$ models as an exact solution in the large-$N$ limit, see e.g.~\cite{Schnitzer:1974ji}.}
\beq
\eqlab{bubble}
T (s) = \frac{1}{\la^{-1} -  (4\pi)^{-2} B(s)}, 
\eeq
where $B(s)$ is a subtracted Passarino-Veltman one-loop integral $B_0$ \cite{Passarino:1978jh}:
\bea
B(s)& \equiv & B_0(s,m^2,m^2)   - B_0(4m^2,m^2,m^2)  =
-2v \,\mathrm{arctanh}\,  v^{-1},
\eqlab{Bint}
\eea
with $m$ denoting the particle mass and $v=\sqrt{1-4m^2/s}$ their relative
velocity. The subtraction is chosen such that at the threshold (zero velocity)
the interaction strength is given by $\la$.
Then, the scattering length is $\ra = - \la/(16\pi m) $ and hence the sign of the potential unambiguously implies that negative or positive $\ra$ corresponds respectively to repulsive or attractive interaction. 

In the center-of-mass frame, the two scatterers share the energy equally and hence their relative momentum is 
\beq
\rk = \half v s^{1/2} = \big(\quarter s -m^2\big)^{1/2}.
\eeq
In the following we use $s$, $v$, or $\rk$ interchangeably as
the energy variable. The amplitude is independent of scattering
angle in this case, hence has no partial waves beyond the $s$-wave.

\begin{figure}
\centering
  \includegraphics[width=9cm]{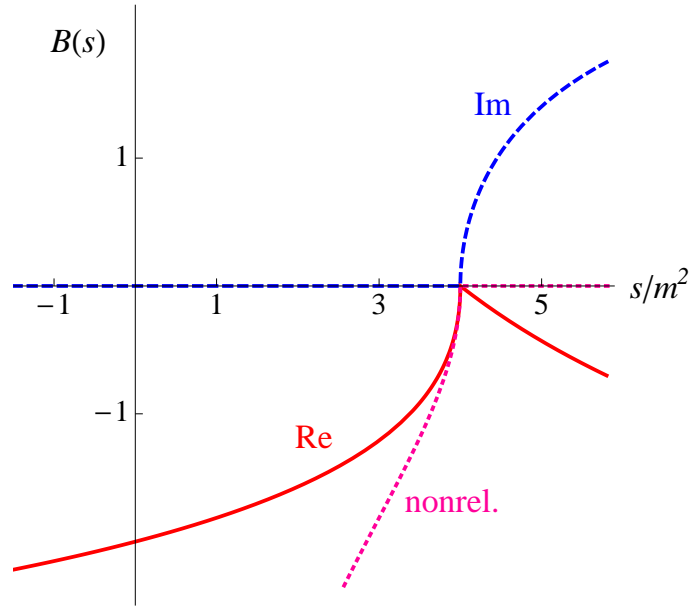}\\
  \caption{The real (solid red) and imaginary (blue dashed) parts of the loop function $B(s)$. The ``nonrel." (dotted magenta) curve shows the non-relativistic approximation to the real
  part.}
    \figlab{Bofs}
\end{figure}

The analytic properties of the amplitude $T$ are determined by the loop
function $B$ plotted in \Figref{Bofs}. For negative $\la$, the amplitude develops a pole at the position where
\beq
(4\pi)^2 \la^{-1} = B(s).
\eqlab{pole}
\eeq
Solving this equation for $s$ one finds the mass squared $M^2$ of the corresponding bound state solution. Since $B(s)$ is negative, there is
no solution for positive $\la$. Furthermore,
above the threshold the loop function develops an imaginary part,
\beq
\eqlab{imB}
\mathrm{Im} \,  B(s) = \pi v \, \theta(v^2) = \frac{\pi }{(1+m^2/\rk^2)^{1/2}}
\, \theta(\rk^2),
\eeq
and since $\la$ is real, there is only a solution below the threshold:
a bound state with  $M^2 < 4m^2$. There are no poles for
complex $s$ as is demonstrated in the Appendix.

Importantly, since the loop function extends to negative $s$,
for $(4\pi)^2 \la^{-1} < -2$ one finds $M^2<0$, i.e.~the tachyon. The appearance of a tachyon solution
is in apparent conflict with causality, and indeed the light-by-light scattering sum rule cannot be satisfied in this case~\cite{Pauk:2013hxa}.
In the bound-state case (i.e., $M^2\geq 0$) the sum rule is satisfied provided
the bound state is treated as an asymptotic state, and hence the channel of $\ga\ga$ fusion into the bound state is included. 

To summarize, while the helicity-difference sum rule given in \Eqref{HDsr} is easily
verified for the repulsive ($\la > 0$) $\de$-function potential \cite{Pauk:2013hxa}, for attractive interaction 
($\la<0$) there is a causal (bound-state) and acausal (tachyon) regimes. We thus distinguish the following two domains:
\begin{subequations}
\bea
\mbox{causal:} && \,  -\infty < \la \leq - 8\pi^2 \,\cup \, \la \geq 0, \eqlab{causal}\\
\mbox{acausal:} && \, - 8\pi^2 < \la < 0.
\eqlab{acausal}
\eea
\end{subequations}
We next consider how these domains project onto the effective-range 
parameters. 

\section{Causality bound in effective-range expansion}
Our suitably normalized elastic scattering amplitude is given by:
\beq
\frac{v}{16\pi} \, T(s) \equiv  f(\rk) = \left( \frac{16\pi}{\la \sqrt{1+m^2/\rk^2} }
 + \frac{2}{\pi} \arctanh \frac{1}{\sqrt{1+m^2/\rk^2 } }\,- i \right)^{-1},
\eeq
and is related to the phase shift
and the $K$-matrix as:
\beq
f(\rk)= e^{i\de(\rk)} \sin \de(\rk) = \big[K^{-1}(\rk) - i \big]^{-1} .
\eeq 
The effective-range expansion proceeds then as follows:\footnote{In doing the expansion one uses (for $x\geq 0$): 
$$ \arccoth \sqrt{1+1/x^2}= \arccosh \sqrt{1+x^2}=\arcsinh x = \sum_{n=1}^\infty 
\frac{(-1)^{n+1} \Gamma(n-1/2) }{(n-1)! (2n-1) \sqrt{\pi} } \, x^{2n-1} . $$}
\bea
\rk \cot \de(\rk)  &= &16\pi\la^{-1} \sqrt{m^2+\rk^2 }
 + (2\rk/\pi) \arccoth \sqrt{1+m^2/\rk^2 }  \nn\\
 &=& 16\pi \la^{-1} m  +  \sum_{n=1}^\infty \frac{(-1)^{n+1} 
 \Gamma(n-1/2)}{(n-1)!(2n-1) \pi^{3/2} } \left( 1+\frac{2n-1}{4n} (4\pi)^2 \la^{-1}\right) 
 \frac{\rk^{2n}  }{m^{2n-1}}.
\eea
Comparing to \Eqref{ERE} we identify
the scattering length, the effective-range and shape parameters:
\begin{subequations}
\bea
\ra &=& -\frac{\la}{16\pi m} ,\\
\rr_1 &= & \frac{4}{ \pi m} \left(1+ 4\pi^2 \la^{-1}\right),\\
\rr_n &= & \frac{2 \, \Gamma(n-1/2)}{n! \, \pi^{3/2} m^{2n-1} }  \left(\frac{n}{2n-1} + 4\pi^2 \la^{-1}\right).
\eea
\end{subequations}
It is obvious that $\rr_n$ can only turn negative provided $\la$ satisfies:
\beq
- 4\pi^2 \frac{2n-1}{n}  < \la < 0.
\eeq
This domain, however,  is well within the acausal region [\Eqref{acausal}],
at least for any integer $n$. Hence, as long as $\la$ is within the allowed causal range
[\Eqref{causal}], we obtain the central result of this work: 
\beq
\rr_n \geq \frac{\Gamma(n-1/2)}{n! \,(2n-1)} \frac{1}{ \pi^{3/2} m^{2n-1} }\geq 0\, ,
\eeq
for any integer $n$. In particular, for the effective range we obtain:
\beq
\rr_1 \geq \frac{2}{ \pi m}.
\eeq
As noted above, this is in near perfect disagreement with the corresponding Wigner's bound: $\rr_1 \leq 0 $.
In the following section we point out a possible origin of this disagreement
and further discuss the analytical properties of the new solution.

\section{K-matrix pole as satellite of the bound-state pole}

The Wigner's bound arises in non-relativistic scattering theory
\cite{Nussenzveig:1972}. Our causality criterion is based on relativistic
dispersion theory. The difference between the bounds
[Eqs.~\eref{Wignerbound} vs.~\eref{newbound}] should therefore be 
pinned on ``relativistic effects". Indeed,
by taking the non-relativistic limit ($\rk/m\to 0$)
in our example one obtains $r_1=0$, which honors 
the Wigner's bound, albeit quite trivially. 

The non-relativistic
limit on the other hand ruins the analyticity in $s$ as
can be seen for  the  ``nonrel." curve in \Figref{Bofs} which
displays the real part of the loop function in the non-relativistic limit; the imaginary part remains unchanged. 
One sees that, while in the threshold region ($s\approx 4m^2$)
the non-relativistic limit may serve as a good approximation, it is missing important features
away from the threshold. One such feature is the  $K$-matrix pole
which appears in relativistic theory at $s=s_K> 4m^2$ such that
\beq
\re\, B(s_K) = (4\pi)^2 \la^{-1}.
\eeq
This pole disappears in the non-relativistic limit, since then 
$\re\, B(s)=0$, for $s\geq 4m^2$. 

In the full theory, however, 
the bound-state pole appearing at $s=M^2<4m^2$ is always accompanied by a $K$-matrix pole.\footnote{The $K$-matrix pole is sometimes
 indicative of a resonance, however not in this
 case.  The present solution for the amplitude $T$, and hence for the
 $S$-matrix, has no poles for complex $s$.  A simple proof of this statement 
 is given in the Appendix.}
 The closer is the bound state to the threshold
(the ``shallower" it is), the closer is the $K$-matrix pole. The phase shift, of course,
crosses 90 degrees at the $K$-matrix pole position, as $K(\rk) = \tan \de(\rk) $ by definition.
Hence, for a shallow
bound state such the deuteron, the corresponding phase-shift (i.e., ${}^3S_1$
in case of $NN$ scattering) 
starts at $\de(0) = \pi$ (due to Levinson's theorem)
and then quickly goes down to cross $\pi/2$ at a fairly low $\rk$. 
This is how in fact the empirical  ${}^3S_1$ phase shift behaves.
In the non-relativistic description with zero-range potential the phase shift
never crosses $\pi/2$. We therefore expect a more effective
description of the deuteron phase-shift within the relativistic theory.

For a very shallow bound state ($\la<0$, $|\la| \gg 8\pi^2$), the transcendental
equations for the bound-state and $K$-matrix pole positions can be solved to yield, respectively:
\begin{subequations}
\bea
M^2 &\simeq & \frac{4m^2}{1+(16\pi)^2 \la^{-2}}, \\
 s_{K} &\simeq & \frac{4m^2}{1+8\pi^2 \la^{-1}} .
 \eea
 \end{subequations}
 We thus can establish an approximate relation between the binding energy,
 $\rB = 2m-M$, and the momentum at which the corresponding phase shift crosses 90 degrees,
 $\rk_{\pi/2} = (1/2) \sqrt{s_K - 4m^2}\,$:
 \beq
 \eqlab{relation}
 \rk_{\pi/2} \approx  \rB^{1/4}\,  m^{3/4}.
 \eeq
 For the kinetic energy $\rk_{\pi/2}^2/m$, we simply have 
 $\sqrt{m \rB}$, which shows that the position of
 the $K$-matrix pole is directly related with the soft scale 
 emerging in the presence of the bound state. This scale arises
 here naturally, rather than as a result of fine-tuning the subleading contributions as in the non-relativistic theory (see e.g., \cite{vanKolck:1999mw}).

\section{Conclusion}
\label{sec4}
The zero-range forces should be playing the leading role in a low-energy EFT
description of any short-range interaction such as nuclear or inter-atomic. 
However, at least in a non-relativistic formulation, a zero-range force is
bound to yield non-positive effective-range parameters~\cite{Phillips:1996ae}, 
and hence is bound not to be adequate empirically, unless a physical
cut-off is introduced.  We have shown that 
in relativistic theory the zero-range force yields only positive effective-range parameters, provided causality is respected.
This appears to be in complete disagreement with Wigner's causality bound.
The precise origin of this paradox has not been entirely understood here, 
however we certainly favor here the relativistic approach to causality. 

A question of consistency of the bubble-chain approximation arises, as from field-theoretic point of view it presents a dramatic truncation
of the full theory. Similar concerns may arise in developing 
a power counting in the EFT framework, as relativistic effects appear merely as effects of ``higher order". The truncation considered in this work is consistent  at least with respect to the agreement with the sum rule, hence
has  the correct analytic structure.

An interesting prediction of relativistic
theory of zero-range interactions is the fact that a bound state is accompanied 
by a $K$-matrix pole. The latter shows up in the pertinent phase-shift 
crossing of 90 degrees. In the case of a shallow bound state, its binding energy
determines the position of the 90 degree crossing
according to \Eqref{relation}. The $K$-matrix pole does not correspond to
a resonance in this case.
 
It remains to be seen whether these findings will help to reorganize the EFT
of nuclear forces so as to defer the finite-range considerations (e.g.,
the pion exchange) and 3-nucleon forces where
the naive dimensional analysis places them --- subleading orders. As result, the idea of `perturbative pions' \cite{Kaplan:1998we},
which fails in the strictly nonrelativistic description, 
may be revived in the relativistic framework.

\section*{Acknowledgements}
It is a pleasure to thank Mike Birse, Evgeny Epelbaum, Ubirajara van Kolck, Dean Lee,  Daniel Phillips, and Marc Vanderhaeghen for valuable remarks on the  manuscript.
The work was partially supported by the Deutsche Forschungsgemeinschaft 
through Collaborative Research Center ``The Low-Energy Frontier of the Standard
Model" (SFB 1044) and
 the Cluster of Excellence ``Precision Physics, Fundamental Interactions and Structure of Matter" (PRISMA).

\appendix
\section*{Appendix: No poles for complex $s$}
To show that the amplitude $T(s)$ given by \Eqref{bubble} has no poles for complex $s$
we need to show that $\la^{-1} = (4\pi)^{-2} B(s)$ has no solution for $s=s_{\mathrm{r} } + i s_{\mathrm{i} } $, with $s_{\mathrm{r} }, \,  s_{\mathrm{i} } \in \mathbb{R}$ and 
$s_{\mathrm{i} } \neq 0$. As due to hermiticity $\la$ is real, we only need to show that
$\im\, B(s)\neq 0$, for $s_{\mathrm{i} } \neq 0$. For this we use the dispersion
relation for the subtracted loop integral: 
 \beq
 B(s)= \frac{1}{\pi} \int\limits_{4m^2}^\infty ds' \, \frac{\im\, B(s')}{s'-s}
 \left(\frac{s-4m^2}{s'-4m^2}\right),
 \eeq
 with $\im \, B(s)$ for real $s$ given in \Eqref{imB}. We then proceed to write
 \beq
  B(s) =\frac{s-4m^2}{\pi} \int\limits_{4m^2}^\infty ds' \, \frac{\im\, B(s')}{|s'-s|^2}\frac{s'-s^\ast}{s'-4m^2}\,.
 \eeq
Hence, the real and imaginary parts of $B$ are given respectively as:
 \bea
 \re \, B(s) &=& \frac{1 }{\pi} \int\limits_{4m^2}^\infty ds' \, \frac{\im\, B(s')}{|s'-s|^2}
 \left( s_{\mathrm{r}} -4m^2+ \frac{|s-4m^2|^2}{s'-4m^2}  \right),\\
 \im \, B(s) &=& \frac{s_{\mathrm{i}} }{\pi} \int\limits_{4m^2}^\infty ds' \, \frac{\im\, B(s')}{|s'-s|^2}.
 \eea
 The integrand in the latter expression is positive definite, hence the integral is not zero, and
 hence for $s_\mathrm{i}\neq 0 $, we indeed have $\im \, B(s)\neq 0$. Therefore,
 $T(s)$ has no poles away from the real axis.

\end{document}